\RequirePackage{fix-cm}
\documentclass[twocolumn,epjc3]{svjour3}  
\journalname{Eur. Phys. J. C}
\usepackage{hyperref}
\begin{document}

\title{Machine Learning is Good for Physics --- and Vice Versa
}

\author{Michael Kr\"amer\thanksref{e1,addr1}
        \and
        Tilman Plehn\thanksref{e2,addr2,addr3} 
}

\thankstext{e1}{e-mail: mkraemer@physik.rwth-aachen.de}
\thankstext{e2}{e-mail: plehn@uni-heidelberg.de}

\institute{Institute for Theoretical Particle Physics and Cosmology, RWTH Aachen University, Germany\label{addr1}
           \and
           Institut f\"ur Theoretische Physik, Universit\"at Heidelberg, Germany\label{addr2}
           \and
           Interdisciplinary Center for Scientific Computing (IWR), Universit\"at Heidelberg, Germany\label{addr3}
}

\date{Received: date / Accepted: date}

\maketitle

\begin{abstract}
Scientific AI is rapidly transforming fundamental physics research and challenging defining aspects of the fundamental physics methodology. We discuss opportunities and dangers of this transformation and find exciting benefits from a close interaction between AI and fundamental physics, provided that we remain aware of the scientific methodologies of the respective fields. For fundamental physics, we discuss two such aspects: statistical validation and a generalizing theory description, both with the goal of discovering new physics in vast datasets.
\end{abstract}

\section{Scientific goal}
\label{sec:goals}

The central goal of fundamental physics research is to identify mathematical structures that describe and relate observations across different physical systems and scales. This is a highly abstract and at the same time quantitative question, where a candidate theory has to describe many measurements and relate different areas of physics, from particle interactions at different energy scales to, for instance, cosmology~\cite{Trotta_2025}.  

Crucially, the goal of fundamental physics is not the description of data per se, but the identification of abstract underlying structures that explain and relate different observations. In particle physics, this program is formulated in a reductionist way, seeking explanations in terms of elementary degrees of freedom and their interactions. Although the observed phenomena can be highly complex, the underlying theory is assumed to become simpler when expressed in terms of the relevant microscopic physics. Over the past decades, fundamental physics has already turned to more complex setups, enabled by numerical analysis techniques and numerical theory simulations. From today's perspective, we should systematically include complex phenomena, changes in the relevant degrees of freedom, non-equilibrium effects, or emergent phenomena, utilizing theoretical predictions in terms of complex numerical simulations.

Historically, this reductionist approach has shaped the development of particle physics experiments towards a preference for relatively simple and well-controlled systems, such as two colliding particles, electron-positron colliders ahead of proton colliders, or a wealth of dedicated experiments targeting specific signals like dark matter interactions. This preference for simplified experimental setups reflects not only human nature, but also an acknowledged difficulty in deriving reliable predictions for complex observations. As a result, the interplay between theoretical control and experimental complexity has been a defining feature of fundamental physics.

Modern observations and datasets are becoming increasingly complex. One example is the precision program at the multi-purpose detectors at the Large Hadron Collider (LHC), another one is 21cm radio astronomy (SKA), both producing and analyzing vast amounts of data. Their link to fundamental physics is provided by theory predictions in the form of first-principles simulations. In particle physics, the comparison of simulated and measured data has a long tradition, nowadays often formalized as simulation-based inference, where theoretical models are tested and constrained through their predictions for observable data. The standard way to deal with complex data is through summary statistics, like low-dimensional histograms or astrophysical power spectra. From this perspective, the central challenge is to work with data representations that retain as much information as possible about the underlying theory parameters.

The promise of modern machine learning (ML) in a fundamental physics context is to provide us with near-optimal representations and inference strategies for these multi-purpose experiments, combining vast amounts of data with theory simulations in a statistically controlled way. This ML perspective may appear technical, but it is conceptually far-reaching. At the same time, it does not change the scientific goals of fundamental physics. It does not exclude further paradigm shifts, but for now it focuses on coping with complexity to answer our deep questions about Nature. Crucially, ML developments do not replace the role of theoretical structures, but extend the set of tools through which they can be connected to data. 

\subsubsection*{Physics discoveries}

Before we describe the relation between AI and fundamental physics, specifically particle physics, in more technical detail, it is useful to clarify the role ML methods play in the methodology of our field~\cite{2024arXiv240518095H,Grosso:2026fgo}. As it is concerned with more than just a description of data, the ontology of particle physics combines data with  concepts such as fundamental particles and symmetries, particle interactions, conserved quantities, temperature, or entropy. The combination of theory and observations is a defining aspect of fundamental physics. 

Physics research and its ontology have always included supporting disciplines, including mathematics and mathematical physics, statistics, numerical methods, and detector engineering. Machine learning should be viewed as a methodological framework supporting different aspects of the research process. ML methods primarily centered around data representations will naturally be most useful for data-driven aspects of fundamental physics. Modern representation learning raises the interesting question whether learned representations in network-defined latent spaces can be related to established physics concepts such as symmetries or effective theories. One may also ask to what extent the structures learned by ML models overlap with the conceptual structures used in fundamental physics. Crucially, ML does not replace the theoretical and statistical frameworks within which physics results are interpreted and validated, but it extends the set of tools available for connecting theoretical models and experimental observations. 

The relation between theory, data, and interpretation also shapes what particle physics considers scientific progress or discovery in the sense of the field's epistemology~\cite{daedalus}. Beyond statistically rigorous  agreement with the data, acceptable theories are expected to satisfy broader conceptual criteria, including mathematical consistency, simplicity, the ability to generalize across different physical systems and scales, and the ability to produce experimentally testable predictions, broadly defined. Historically, a leading theme has been reductionism, starting from simplified experimental setups and gradually extending the theoretical description to increasingly complex phenomena. Because of its quantum nature, particle physics relies on statistical inference rather than deterministic validation. As a result, a rigorous statistical framework is a defining part of the field. This affects in particular the notion of \textsl{discovery}: an experimental observation has to be statistically validated and, at least in principle, connected to a broader theoretical description. Discoveries made in one experiment or at one energy scale have to be testable in different experimental settings or in different areas of fundamental physics.\footnote{Physics history contains many presumed `discoveries', which eventually disappeared once additional data or improved statistical analyses became available.} 

These ontological and epistemic considerations lead us to clearly separate the two directions \textsl{ML for Physics} and \textsl{Physics for ML}. We focus on the former, where ML by itself does, for now, not affect the scientific objectives of fundamental physics. There are cases where a scientific result implies progress in both directions, naturally leading to both a physics and an ML interpretation of the same work. However, physics applications are always evaluated according to the scientific standards of physics. For example, perturbative quantum field theory remains extraordinarily successful despite its formal limitations because it provides quantitatively accurate and widely generalizable descriptions of data. Another example is numerical algorithms in physics, which do not necessarily require mathematically unique solutions, as long as the resulting predictions are physically meaningful and experimentally validated. 

Finally, scientific methodologies are not immutable. Research fields evolve with time, new data, new technologies, and changing theoretical perspectives~\cite{feyerabend}. Scientific communities can develop preferred ways of formulating questions and interpreting results, sometimes in productive ways and sometimes in ways that only become apparent in hindsight. In this sense, the role of abstraction and theory in increasingly data-driven sciences may also evolve. However, such changes should emerge from the development of fundamental physics itself rather than from adopting ML methodology uncritically as an external framework.

\section{AI in particle physics}

Given our background, we focus on particle physics, assuming or at least hoping that our comments generalize to other areas of fundamental physics. From a theory perspective, there are good reasons to expect that this generalization holds, such that particle physics data and cosmological observations are eventually described by quantum field theory~\cite{Srednicki:2007qs,Schwartz:2014sze}. This underlying theory is encoded in a common Lagrangian or an action. Given a particle content and a set of symmetries, in particular local gauge symmetries, the Lagrangian determines all interactions. The ultimate goal of discovering new physics can then be phrased as extracting the Lagrangian that describes all particle physics or even fundamental physics data. One straightforward example is a dark matter agent with a given mass and set of interactions. Another one is the explanation of the matter-antimatter asymmetry through the Sakharov conditions, including CP violation.

The question whether this Lagrangian is fundamental in the ultraviolet or should be understood as an effective field theory is reflected in the dimensionality of its interaction terms. Renormalizable theories and their Lagrangians can be evolved across energy scales through renormalization group running. Transitions between different sets of propagating degrees of freedom, such as bound states from strong interactions, are included within this framework. While there might eventually be reasons to extend or even abandon this quantum field theory framework, we do not expect ML methods by themselves to trigger such a drastic paradigm shift.

Predictions from quantum field theory employ symbolic as well as numerical methods. Precision hadron collider physics aims to describe as many measurements at as many energy scales as possible. While high-energy predictions for the LHC are based on perturbative quantum field theory, first-principle predictions at lower energies require non-perturbative methods, including lattice gauge theory, to describe the observed hadrons and their spectra. Generative ML accelerates and transforms theory simulations, not at the cost of their first-principles nature, but enabling first-principles predictions at levels of complexity where we are severely limited by numerics or where we traditionally resort to ad-hoc modeling. In terms of AI concepts, this program can be viewed as generating digital twins of the observed data, predicted by first-principles theory, analyzed using simulation-based inference~\cite{Cranmer:2019eaq}.

Particle physics analyses then rely on two established strategies. First, searches for new particles or interactions define a hypothetical beyond-the-Standard-Model (BSM) Lagrangian and test this hypothesis in a given final state or signature. The standard approach is a hypothesis test between the Standard Model (SM) and the BSM model, both implemented through theory simulations, with the Neyman--Pearson lemma providing the statistical foundation. Such a hypothesis test is formally not required for a statistical discovery, but it ensures that searches using a hypothesis test automatically come with a theory interpretation as part of the simulation.

Second, precision measurements typically target Lagrangian parameters, such as masses or couplings. Historically, precision measurements assumed a fixed set of propagating particles, but effective field theory interpretations based on Wilson coefficients soften this assumption. All these model parameters are measured as precisely as possible within a fiducial phase space region. The statistical framework for continuous parameter inference is based on information geometry, including the Cramér--Rao bound. 

Recently, both strategies have been combined, for example in precision measurements of differential cross sections in a fiducial phase space, followed by subsequent interpretations in terms of Standard Model parameter extraction, Standard Model Effective Field Theory (SMEFT) parameter extraction, or BSM searches~\cite{ATLAS:2025oiy}.

\subsection{ML-enhanced classic analyses}

Essentially, every aspect of a classic particle physics analysis, including both searches and precision measurements, is being transformed and enhanced through ML methods. Triggering will become more efficient and flexible through ultra-fast neural networks evaluated in real time. Data acquisition and the combination of detector information will make use of learned latent representations. Object identification and classification were the starting point of deep networks in particle physics~\cite{deOliveira:2015xxd} and have become standard examples of ML outperforming established methods~\cite{Komiske:2018cqr,Qu:2019gqs,Kasieczka:2019dbj}. Detector and object calibration are natural targets for ML improvements. In all of these cases, ML methods lead to quantifiable and significant improvements in performance and computational speed and efficiency. Ideally, these improvements can be realized without loss of control and understanding, depending on the methods we develop and employ.

On the theory side, ML-methods can simplify analytic expressions, solve differential equations, or provide optimal implementation of standard integrals. First-principle event generators, based on perturbative quantum field theory~\cite{Campbell:2022qmc}, still include a limited set of tuning parameters, such as factorization, renormalization, and shower starting scales. Parton showers are described by large logarithms in quantum field theory and implemented as (possibly enhanced) Markov processes. Hadronization or fragmentation rely on phenomenological models with a larger set of tuning parameters. Hadron decays are based on parametrizations inspired by quantum field theory and constrained by measured hadron properties. For every step in this simulation chain we search for improvements from broadly defined generative ML. A conservative high-impact strategy targets numerical implementations of established physics methods and applies ML to remove performance bottlenecks and enable higher-precision simulations~\cite{Butter:2022rso,Janssen:2025zke,DeCrescenzo:2026tsp}. 

In ML-enhanced analyses, AI mostly implies neural networks for regression, classification, and (conditional) generation. From a particle physics perspective, these neural networks can be understood as flexible fits or density estimators. The actual LHC analysis is a statistical task, comparing simulated and measured events. Providing the necessary simulations, collecting and calibrating the data, and performing the statistical analysis will involve ML methods at every stage. These will not pose any problem as long as the ML tools are properly controlled and not used as an excuse for statistical shortcuts.

\subsection{ML-enabled analyses}

In addition to enhancing and accelerating classic analyses, ML will also transform the way we analyze data. Since the future is notoriously hard to predict, we start with two historical examples from the LHC. One example is the development of the LHC into the first precision hadron collider, something that was conceptually not foreseen and not part of the original case for the LHC. Another example is the treatment of hadronic final states. When the LHC program was conceived, jets were primarily used as inclusive observables. In modern LHC analyses, they are understood in much greater detail and exploited as structured objects, bridging event-level and subjet-level precision physics. These examples are not meant to anticipate the impact of ML, but to illustrate how conceptual transformations in data analysis have always driven progress in fundamental physics.

One challenge we are facing in fundamental physics is the rapidly increasing size of scientific datasets. A candidate for ML-inspired analysis techniques targeting this challenge is anomaly searches. Their weakly supervised variants generalize so-called bump hunts, where a histogrammed kinematic observable is tested by fitting a function to all points except for a sliding-window signal region. If the quality of the fit improves for a certain position and size of the window, the data indicate an unexplained structure. Weakly supervised anomaly detection extends this idea by learning a background model for potential signal regions with a neural network~\cite{Nachman:2020lpy,Hallin:2021wme}. This example illustrates several common aspects of new ML-based analysis techniques: (i) they are often not entirely new but build on particularly challenging existing concepts; (ii) without ML, the classical methods were not practically viable; (iii) the main challenge lies in the statistical analysis, from controlling the look-elsewhere effect~\cite{Hein:2025ysv} to understanding the systematics of the network training.

The next step in anomaly searches employs learned anomaly scores, for instance, provided by (normalized) autoencoders~\cite{Heimel:2018mkt,Farina:2018fyg,Dillon:2022mkq}. These methods are well established in the ML community and will be useful, for example, to monitor detector performance or improve trigger strategies. For physics analyses they will only become fully useful once embedded in a consistent statistical framework. 

A second, general direction of new ML-analysis techniques is analysis reinterpretation, with the technical aspect of deconvolution or unfolding~\cite{Andreassen:2019cjw,Bellagente:2020piv}. The basic idea is not new: radio astronomy data are recorded as interference patterns in frequency space, Fourier-transformed to physical space, and often analyzed in terms of power spectra. At the LHC, unfolding to kinematic observables is routinely performed. However, classical methods are limited through their poor scaling with the dimensionality, and binned histograms are far from optimal. Interest in ML approaches is driven by their potential to turn reinterpretation into a standard method for making experimental data publicly available. 

\subsection{Representation-based analyses}

The examples in the previous section lead us to a general concept behind ML-enabled analyses, namely the transition from predefined feature spaces to learned latent representations. Many ML applications in particle physics can be understood as representation learning~\cite{DBLP:journals/corr/abs-1206-5538}. Rather than relying on predefined observables, these methods learn internal descriptions of the data optimized for a given analysis or inference task. The central idea is that analyses benefit from an optimal representation of the data. In particle physics we know the underlying symmetry structure of the data, which means that an optimal representation will benefit from explicitly equivariant or covariant architectures~\cite{Brehmer:2024yqw,Favaro:2025pgz,DBLP:journals/corr/abs-2104-13478} as well as implicitly learned symmetries and structures \cite{Golling:2024abg,Birk:2024knn,Mikuni:2024qsr}. The fact that in particle physics symmetries are typically broken softens this contrast to the point where we can either learn the breaking of explicit symmetries or the broken symmetries from scratch.

Moreover, learned representations may combine information from different data modalities, be it different detector outputs in particle physics or different telescopes in multi-messenger cosmology. Modern network architectures, including foundation models~\cite{DBLP:journals/corr/abs-2108-07258}, provide powerful new tools for constructing such representations.

The use of optimized data representations is not new in particle physics. Much of particle physics can be viewed as a search for representations that capture the relevant physical information at minimal complexity. Particle flow, which combines detector information into an optimal abstract description, has long been a driver of the LHC program. On the theory side, jets defined through suitable reconstruction algorithms have long served as standard analysis objects. More recently, this paradigm has been extended to jet substructure, providing access to increasingly fine-grained information. In this sense, learned latent representations are a powerful extension of existing ideas.

A crucial difference between learned representations and traditional representations in particle physics is that the latter are defined in terms of an underlying theory. From a more formal perspective, the particle physics simulation chain can be interpreted as a sequence of factorized conditional probabilities, starting from the low-dimensional hard scattering process defined by the Lagrangian parameters. Each step in this chain defines a representation of the underlying physics, some more directly interpretable than others. Even the particles entering the detector are only probabilistically related to reconstructed objects, and their interpretation involves model assumptions, for example when attributing missing transverse momentum to neutrinos rather than new dark matter particles. The notion of representation in particle physics is inherently theory-informed, and learned representations should ultimately be connected to this structure. The challenge is therefore not merely to learn useful representations, but to understand how they relate to the theory-driven representations already used throughout fundamental physics. 

How to extract quantitative fundamental knowledge from learned latent representations without loosening the scientific standards of a field like particle physics is an open and exciting problem. It raises the hypothetical question of what would happen if ML methods were to identify a previously unknown pattern or structure in the latent representation of the LHC dataset. Such a structure would not by itself constitute a fundamental physics discovery. A fundamental physics discovery has to meet the domain-specific statistical requirements and should eventually lead to a generalizing theory prediction. An illustrative example would be a discovery of extremely weakly interacting particles in semi-visible jets triggered by a signal in an ML-anomaly search at the LHC. The proper discovery has to include a statistical analysis with an understood background model and will be accompanied by quantum field theory work identifying a signal model with a set of fundamental particles and interactions. This signal model should generalize and can lead to the claim of a discovery of dark matter provided it describes the production mechanism during the evolution of the universe and shows that a long lifetime on LHC time scales means a long lifetime on cosmological time scales.

\subsection{Agentic physics research}

In particle physics, an acceleration of experimental and theoretical workflows is not only welcome but increasingly necessary. Precision predictions and precision analyses often require time scales that exceed the lengths of typical PhD or postdoctoral positions. These time scales are not simply a feature of our field, but represent a serious structural problem. The question is therefore not whether parts of the coding and workflow should be automated, but which parts can be delegated reliably to computational systems. Agentic approaches can help accelerate many aspects of our workflows. Unlike standard ML applications, which provide tools for specific tasks, agentic systems orchestrate sequences of tasks, combining data access, simulation, statistical inference, and interpretation. In this way, they change how established analyses are constructed, executed, and communicated.

We can distinguish between two types of agentic systems. The first type consists of general-purpose LLM agents without access to domain-specific tools. Such agents can, for example, compare datasets, perform standard statistical tests, or coordinate general data-analysis tasks. While useful, these capabilities remain at the level of general data analysis and are currently not sufficient to extract physics knowledge. An exciting open question is whether LLM-based agentic systems will eventually generate meaningful physics knowledge on their own, or whether LLMs are, by themselves, a suitable architecture for the quantitative sciences. The same question arises, for instance, in materials design.

The second type consists of LLM agents that interact with domain-specific tools, such as event generators, detector simulations, likelihood evaluations, or analysis frameworks~\cite{Diefenbacher:2025zzn,Plehn:2026gxv}. These tool-augmented agents can navigate substantial parts of the analysis chain, from generating hypotheses and producing simulated data to comparing predictions with measurements and updating model parameters. The development of such systems requires well-defined interfaces to existing numerical tools and a consistent integration of physics knowledge.

Moreover, agentic systems for particle physics should be understood as interactive systems that can be queried, guided, and corrected. An agent can be asked to perform a sequence of analysis steps, explain intermediate results, or explore alternative assumptions. This type of interaction goes beyond traditional analyses, where much of the detailed knowledge is encoded in complex internal workflows and is often difficult to reconstruct once an analysis is completed. This perspective also connects to future scientific communication, in which agents may both produce and consume research output. In particle physics, related structures already exist: large LHC collaborations summarize highly complex analyses in publications, while the underlying data, software, and documentation are only partly accessible outside the collaboration. Agentic systems will make such workflows more transparent by providing interactive access to the underlying analysis workflow, including intermediate results and computational procedures. This opens an especially promising direction of agentic physics, namely re-simulations for existing analyses and analysis re-casting in view of new theory hypotheses.

\section{Physics challenges to AI methods}

Particle physics not only applies ML methods, but also poses questions that go beyond the usual ML scope, similar to its historical role in detector and computing technologies. These questions arise because fundamental physics places unusually stringent requirements on statistical inference, uncertainty quantification, and theory interpretation. These questions can, but do not have to inspire research in the direction \textsl{Physics for ML}.

First, particle physics views classification as learning class probabilities~\cite{Metodiev:2017vrx}. Learning continuous probability densities from training instances with discrete labels requires the same density estimation step as generative models trained on weighted or unweighted events. The learned class probabilities are combined with rate predictions to obtain a predicted production rate, requiring a calibrated uncertainty estimate on the class probability. These requirements are uncommon in standard ML benchmarks, but arise naturally in particle physics applications. Similarly, a typical simulation and ML-simulation question asked in particle physics is \textsl{If a generative network is trained on a million events, how many events can be simulated before being limited by the training statistics?}~\cite{Butter:2020qhk} Questions of this type are less common in ML, but naturally arise in particle physics simulations. They can be answered within the statistical framework developed for particle physics applications.

In particle physics we must ensure that the implicit bias from the data representation, the network architecture, and the hyperparameters does not lead to significant systematics in the learned central values and uncertainties. Avoiding and controlling biases has been a critical motivation of using neural networks rather than fit functions to encode the parton densities for instance in the proton. Their treatment of biases in network design and training goes well beyond what is typically required in ML research~\cite{NNPDF:2024nan}. A related issue concerns biases introduced through learned representations. Foundation models combine different input modalities such that the latent representation can infer information from one modality through correlations with another. These correlations are learned from the training data and can be viewed as an extrapolation in the joint input space. For particle physics applications this requires a reliable uncertainty estimate.

As mentioned above, anomaly searches are an integral and exciting part of the standard ML toolbox, but for a particle physics discovery they ultimately require either an understood background model or a statistical interpretation of the anomaly score. In a similar vein, the acceleration of ML analyses and the combination with anomaly detection raises the question of how to maintain the notion of global vs local significances, accounting for the rapidly increasing number of analyses. Interestingly, these are not themselves fundamental physics questions, but arise from the need to combine a rigorous statistical framework with ML methods in fundamental physics applications. Faced with such a choice, fundamental physics will continue to prioritize its statistical foundations, which have repeatedly protected the field from false discoveries.

Moving to representation learning, the key question is how we can relate such a representation to a well-defined theory. The first question is how a learned representation quantitatively encodes the information in the training data. Information geometry provides a framework for analyzing latent spaces, allowing concepts such as geodesic distances, curvature, and non-metricity to be used for quantitative explainability~\cite{arvanitidis2021latentspaceodditycurvature,Kuntz:2026kuv,Bal:2026pzw}. This illustrates how concepts originating in fundamental physics can contribute to quantitative explainability in ML. Combining physics methods with ML also leads to the question of how we can understand learning patterns using established concepts from complex systems physics~\cite{Roberts:2021fes,ringel2025applicationsstatisticalfieldtheory}. Finally, while in an ML context there exists a clear separation between implicitly and explicitly encoding symmetries in a network architecture, the symmetries relevant for experimental particle physics are typically broken by the experimental setup. This resolves the strict division between an explicit vs implicit treatment of symmetries to the question if we want to learn the breaking of a known symmetry or the broken symmetry from scratch, and which way we extract the most knowledge.

\section{Dangers and opportunities}

There are good reasons to critically discuss the impact of AI on fundamental physics research. The question is not whether we will use AI methods, but how we integrate them into the scientific standards of our field. In particular, AI should not weaken the two pillars of fundamental physics: controlled statistical inference and generalizing theory interpretation.

\subsubsection*{Potential dangers}

The first danger is that ML methods are used outside a proper statistical framework. Fundamental physics has developed very high standards for comparing experimental data with theory predictions, including uncertainty estimates, large numbers of nuisance parameters, and hypothesis tests. There is no reason to relax these standards just because ML methods compare data and predictions in high-dimensional spaces instead of summary observables. On the contrary, the more powerful and flexible the method, the more important it becomes to control its statistical interpretation. AI-assisted analyses therefore have to remain embedded in a proper statistical framework with clearly defined discovery criteria.

A second danger is the replacement of first-principles theory predictions by ML-defined `theory' models. Such models may describe existing data very well, which can be useful within a given analysis. However, describing data successfully is not the same as providing a theory of fundamental physics. A theory should not only reproduce observations, but also provide an abstract structure which explains them and generalizes. In particle physics, this structure is given by quantum field theory. Without such a framework, an ML model of proton-proton collisions would have limited explanatory power and would not allow us to connect LHC measurements to nuclear physics, flavor physics, or cosmology.

A third, more practical danger is an inflation of AI-accelerated analyses and publications with limited novelty. At the LHC it will become increasingly easy to run many weakly supervised anomaly searches, reinterpretations, or high-dimensional scans on existing and new data. Such analyses can be useful, but the AI-accelerated environment forces us to define more clearly what we consider a scientific contribution: a new physics question, an improved level of control, a new interpretation, a new connection between measurements, or a method that can be reused. Simply applying a powerful workflow to another dataset will often not be sufficient. We note that tightened scientific standards will make it hard for purely agentic and for purely human contributions to be published.

\subsubsection*{Scientific opportunities}

However, our reaction to these dangers cannot just be conservative. AI will change fundamental physics; it is an opportunity, not a problem. Its scientific promise lies in accelerating analyses and calculations, improving the quality of simulations, enabling more powerful and flexible analyses, and making complex workflows more transparent. If we want to benefit from these developments, it is important that we maintain methodological clarity. The goal of fundamental physics is not the automated production of descriptions of data, but the understanding of the underlying structures. AI can help us achieve this goal, but it should not redefine it.

\subsubsection*{University environment}

Most fundamental physics research takes place at universities. There, we have to ask ourselves whether society ultimately values fundamental physics primarily because of discoveries such as the Higgs boson or dark matter, or because it educates highly qualified graduates. We should be open to the view that the societal relevance of fundamental research is not only driven by research results, but also by our training of students and young researchers. A large fraction of them will eventually move into industry or other sectors, where they contribute with their analytical skills, problem-solving strategies, experience with complex data and models, and international experience. In the era of LLMs and AI agents, fundamental physics research should teach young researchers how to ask the right questions and solve them with cutting-edge technology. This educational mission therefore has to become part of our AI strategy.

In addition to scientific considerations, universities also have to consider the environmental impact of AI. In the same way that the CO${}_2$ footprint of the LHC has to be considered during design and funding, the rapidly increasing energy consumption of generative AI tools requires careful justification~\cite{Luccioni_2024}. Physics research, with its focus on performance, should explore whether general AI tools trained and deployed in large-scale AI data centers are always the best solution once environmental and ethical aspects are taken into account. In academia, the AI question is therefore not only about smart versus large networks, but also about responsibility and resource efficiency.

These two university-specific aspects define our program for the future. The methods and values we apply in research will also shape the next generation of scientists. As AI methods become central to data analysis, simulation, and interpretation, they have to become central to our training. This starts with learning how to apply these methods and tools, but also how to understand their limitations, control their uncertainties, optimize their implementation, and embed them into a proper scientific framework. If we fail to incorporate scientifically sound AI methods into our teaching and training, we risk losing scientific relevance and part of our broader societal impact. If we succeed, AI will not only change how we do physics, but also strengthen the role of fundamental physics in a rapidly evolving scientific and technological landscape.

\section{Future fundamental physics}

Scientific AI is transforming all aspects of our lives, including fundamental physics research. The potential gains for physics research are huge, because its reasoning is not based on an abstract learnable language, like mathematics or part of computer science, but driven by understanding increasingly complex real-world data.

A major challenge is to align the scientific AI methodology with the defining elements of fundamental physics, most notably its level of mathematical abstraction, its quantitative understanding, and its rigorous statistical framework. Fundamental physics comes with an established methodology into which AI methods have to be integrated to contribute meaningfully to physics discoveries and knowledge gain. We focus on two aspects, the statistical validation and generalizing theory descriptions.

We have identified four ways in which scientific AI is changing particle physics research. First, essentially all classic analysis steps have been shown to benefit from ML, from engineering-inspired detector read-out to quantum field theory predictions. Second, ML is enabling new analysis ideas, for instance, anomaly detection or analysis re-interpretation. Both, ML-enhanced and ML-enabled analysis techniques are being integrated in the particle physics methodology in view of the upcoming LHC runs. 

Third, representation learning is related to the abstract, quantitative, and generalizing knowledge representation in terms of quantum field theory and its action or Lagrangian. How to combine this knowledge representation with learned latent representation is an exciting open problem. 

Fourth, agents are massively accelerating workflows. Agentic systems working with domain-specific numerical tools are easily included in the scientific method of fundamental physics, whereas agentic systems developing and applying their own tools still have to prove their potential and develop a relation to established research methodologies. In experimental and theoretical particle physics, accelerated research is an entirely positive development. We emphasize that of these four ML directions only the acceleration of scientific workflows  is primarily driven by multi-purpose LLMs. Most physics applications of modern AI are more closely tied to architectures developed for other scientific or engineering applications.

An interesting question is how fundamental physics can, in turn, challenge AI methods and trigger further developments. Again, the leading theme is the requirement to align statistics and ML, both applied to fundamental physics questions. Other promising directions include complex system concepts or differential geometry applied to neural networks or learned latent representations. 

When evaluating the promise of applying scientific AI to fundamental physics we need to remind ourselves that most research is done at universities. Knowledge gain is only one objective, alongside the broader societal impact of university teaching and training. Evaluating fundamental physics solely in terms of knowledge gain will lead to a reduction in societal and funding support. The focus on training young scientists and future leaders implies that we have to embrace scientific AI for fundamental physics research without cutting scientific corners, to remain relevant in science and in society.

Finally, the AI transformation of society and research is happening much faster than historical precedents and leaves little time for scientific disciplines to adapt. Fundamental physics has to embrace AI while maintaining the methodological standards that have shaped the field for decades. If we succeed, AI will become an essential part of our methodology, will help us achieve our scientific goals, and will inspire us to define new scientific goals. 

\subsubsection*{Acknowledgements}

We would like to thank Jesse Thaler for triggering our interest and Lydia Brenner, Stefano Forte, Louis Lyons, Markus Elsing, and David Shih for their helpful comments. We are supported by the Deutsche Forschungsgemeinschaft (DFG, German Research Foundation) under grant 396021762 -- TRR~257 \textsl{Particle Physics Phenomenology after the  Higgs Discovery} and by the KISS (05D23GU4) and SciFM (05D25VH3) consortia funded by BMFTR in the ErUM-Data action plan.

\bibliographystyle{SciPost_bibstyle}
\bibliography{tilman,refs}
\end{document}